\documentclass{fastzero}
\usepackage[nolist]{acronym}

\usepackage{amssymb}
\usepackage{pifont}

\begin{acronym}[XXXXXXXXX]
\acro{CA}{Cooperative Awareness}
\acro{CAV}{Connected Automated Vehicle}
\acro{CAM}{Cooperative Awareness Message}
\acro{CCAM}{Cooperative, Connected, and Automated Mobility}
\acro{CP}{Cooperative Perception}
\acro{CPM}{Cooperative Perception Message}
\acro{CV}{Connected Vehicle}
\acro{CSMA/CA}{Carrier Sense Multiple Access with Collision Avoidance}
\acro{C-ITS}{Cooperative Intelligent Transport System}
\acro{C-V2X}{Cellular Vehicle-to-Everything}
\acro{DCF}{Distributed Coordination Function}
\acro{DDT}{Dynamic Driving Task}
\acro{DEN}{Decentralized Environmental Notification}
\acro{DENM}{Decentralized Environmental Notification Message}
\acro{EC}{European Commission}
\acro{ETSI}{European Telecommunications Standards Institute}
\acro{GCM}{GNSS Correction Message}
\acro{GNM}{Gradient Navigation Model}
\acro{GNSS}{Global Navigation Satellite System}
\acro{GPS}{Global Positioning System}
\acro{GUI}{Graphical User interface}
\acro{HFC}{High-Frequency Container}
\acro{ICV}{Intelligent and Connected Vehicle}
\acro{IGG}{Inter-Generation Gap}
\acro{IPG}{Inter-Packet Gap}
\acro{ITS}{Intelligent Transport System}
\acro{ITS-S}{ITS Station}
\acro{IVIM}{Infrastructure-to-Vehicle Information Messaging}
\acro{LLC}{Logical Link Control}
\acro{LFC}{Low-Frequency Container}
\acro{LSM}{Least Square Method}
\acro{LTE}{Long-Term Evolution}
\acro{MAC}{Medium Access Control}
\acro{MQTT}{MQ Telemetry Transport}
\acro{MSE}{Mean Squared Error}
\acro{OMNeT++}{Object Modular Network Testbed in C++}
\acro{OEDR}{Object and Event Detection and Response}
\acro{OSI}{Open Systems Interconnection}
\acro{NR}{New Radio}
\acro{P2I}{Pedestrian-to-Infrastructure}
\acro{P2V}{Pedestrian-to-Vehicle}
\acro{P2X}{Pedestrian-to-Anything}
\acro{PSM}{Personal Safety Message}
\acro{RRM}{Roadside Ranging Message}
\acro{SAE}{Society of Automotive Engineer}
\acro{SPAT}{Signal Phase and Timing}
\acro{SPS}{Semi-Persistent Scheduling}
\acro{SUMO}{Simulation of Urban Mobility}
\acro{SFM}{Social Force Model}
\acro{TCP}{Transmission Control Protocol}
\acro{UDP}{User Datagram Protocol}
\acro{VA}{Vulnerable Road User Awareness}
\acro{VAM}{Vehicle Awareness Message}
\acro{V2I}{Vehicle-to-Infrastructure}
\acro{V2V}{Vehicle-to-Vehicle}
\acro{V2X}{Vehicle-to-Anything}
\acro{V2P}{Vehicle-to-Pedestrian}
\acro{VAM}{VRU Awareness Message}
\acro{VANET}{Vehicular Ad hoc Network}
\acro{VRU}{Vulnerable Road User}
\end{acronym}

\begin{document}
\twocolumn[
\begin{titlepage}
    \papercategory{Scientific Paper}
    \papertitle{V2X Intention Sharing for Cooperative Electrically Power-Assisted Cycles}
    \authors{Felipe Valle, Johan Elfing, Joel P\aa lsson, Elena Haller, Oscar Amador}
    \affiliations{Halmstad University}
    \contactinfo{Contact address: Halmstad, Halland, 301 18, Sweden \\
    Phone: (+46) 035-16 71 00 \\
    Corresponding author's e-mail: oscar.molina@hh.se}
    \symposiumtopic{A-1: Collision Avoidance; A-6: Connected \& Cooperative Driver Assistance Systems; B-7: Vulnerable Road Users}
\end{titlepage}
]

\thispagestyle{fancy}

\begin{abstract}
    This paper introduces a novel intention-sharing mechanism for Electrically Power-Assisted Cycles (EPACs) within V2X communication frameworks, enhancing the ETSI VRU Awareness Message (VAM) protocol. The method replaces discrete predicted trajectory points with a compact elliptical geographical area representation derived via quadratic polynomial fitting and Least Squares Method (LSM). This approach encodes trajectory predictions with fixed-size data payloads, independent of the number of forecasted points, enabling higher-frequency transmissions and improved network reliability. Simulation results demonstrate superior inter-packet gap (IPG) performance compared to standard ETSI VAMs, particularly under constrained communication conditions. A physical experiment validates the feasibility of real-time deployment on embedded systems. The method supports scalable, low-latency intention sharing, contributing to cooperative perception and enhanced safety for vulnerable road users in connected and automated mobility ecosystems. Finally, we discuss the viability of LSM and open the door to other methods for prediction.
\end{abstract}

\section{Introduction}
\acp{VRU} make up to 20\% of fatalities in road accidents globally~\cite{Who2022}. Initiatives such as Vision Zero~\cite{ecVisionZero}, aimed at reducing deaths and serious injuries to zero by 2050, consider the use of \ac{CCAM} to ensure road safety and traffic efficiency. 

While existing \ac{CCAM} services allow road users to predict or infer each other's movements (i.e., intention detection)~\cite{detection-sick-ml}, true cooperation (the first C in CCAM) implies sharing goals and strategies (i.e., the definition of a cooperative game~\cite{games}). Furthermore, intention sharing reduces and distributes the complexity of tasks compared to intention detection: instead of having $n$ road users compute $n-1$ predicted trajectories, they only predict and share their intended trajectory and receive those of their neighbors.

The latter approach is the one envisioned by the \ac{ETSI} in their \ac{VRU} awareness basic service~\cite{etsiVAM}, which relies on \acp{VAM} including \acp{VRU} position and velocity (speed and heading), and its predicted trajectory as a set of coordinates. However, due to the data requirements to include all this information in every message, VAMs only include this prediction in a subset of VAMs (i.e., in a container that is sent only every 2\,s) and with a prediction limit of up to 10\,s or 40 points ahead.


In this work, we propose an optimization for intention sharing in vehicular networks. Instead of transmitting raw coordinate sequences, the predicted trajectory is encoded as a geometric shape representing a geographical area. This representation preserves the essential information—future positions and associated uncertainty—while significantly reducing the data payload. As a result, whether the prediction comprises 4 or 40 points, the same number of bytes is required to encode the area. This reduction in message size enables more frequent transmissions, thereby improving network reliability and communication efficiency.

To this end, we conduct a comparative analysis of the trajectory representation formats permitted by the standard, with the goal of evaluating their suitability across diverse communication and processing scenarios. In addition, we perform a complexity analysis to quantify the computational and communication trade-offs associated with each representation method. Finally, we assess the proposed strategy through both simulation and real-world implementation.


\section{Related Work}

The problem of maneuver coordination has traditionally been addressed from the perspective of intention detection, as outlined in the ETSI standards \cite{etsiVAM}. In \cite{schulz2015pedestrian} the authors proposed a latent-dynamic conditional random field model for pedestrian intention recognition, focusing on crossing behaviors in urban environments. Their extended work introduced a multiple-model filter for simultaneous intention recognition and trajectory prediction~\cite{schulz2015controlled}. The work done in \cite{quintero2015pedestrian} introduced a model combining dynamics and behavior classification for pedestrian path prediction. In the context of cyclists, Pool et al.~\cite{pool2017road} developed a route prediction method that integrates road topology. For motor vehicles, Benterki et al.~\cite{benterki2019lane} applied machine learning to detect lane-change intentions. 

However, a major limitation of intention detection as defined in the ETSI standard is its reliance on transmitting the last 23 trajectory points per user. This results in substantial communication overhead, especially in dense traffic scenarios. Consequently, there has been significant research focused on optimizing the generation and triggering of VAMs to mitigate this burden~\cite{martínpérez2023, silas2024}. 

Furthermore, the standard attempts at avoiding overhead by only sending crucial information (i.e., position, speed, and heading) in the High-frequency Container that is sent every VAM, while only sending optional containers (e.g., indicating past and predicted paths, clustering operations, and other information about the VRU) at longer intervals~\cite{etsiVAM}. However, this variability in sizes creates problems at the Access layer --- for WiFi communications, it increases the probability of a collision, and in cellular communications, it is problematic for radio resource allocation~\cite{Anupama2022}. Having VAMs of uniform size translates into better performance metrics such as Age of Information (AoI) and PDR~\cite{thesis-IS}.

An alternative to intention detection is intention sharing, wherein vehicles explicitly broadcast their intended maneuvers. This approach offers several notable advantages. First, it minimizes uncertainty propagation, since each agent predicts only its own future trajectory rather than inferring that of others. Second, by decoupling prediction from perception, intention sharing reduces the computational load on receiving vehicles. Third, compared to intention detection schemes—such as those defined in ETSI standards, which require broadcasting full trajectory histories—intention sharing enables more compact encoding (e.g., geometric areas or maneuver primitives), thus significantly lowering communication overhead and improving scalability in high-density traffic.

\section{Trajectory Prediction}

\begin{figure}[tbh!]
    \centering
    \includegraphics[width=\linewidth]{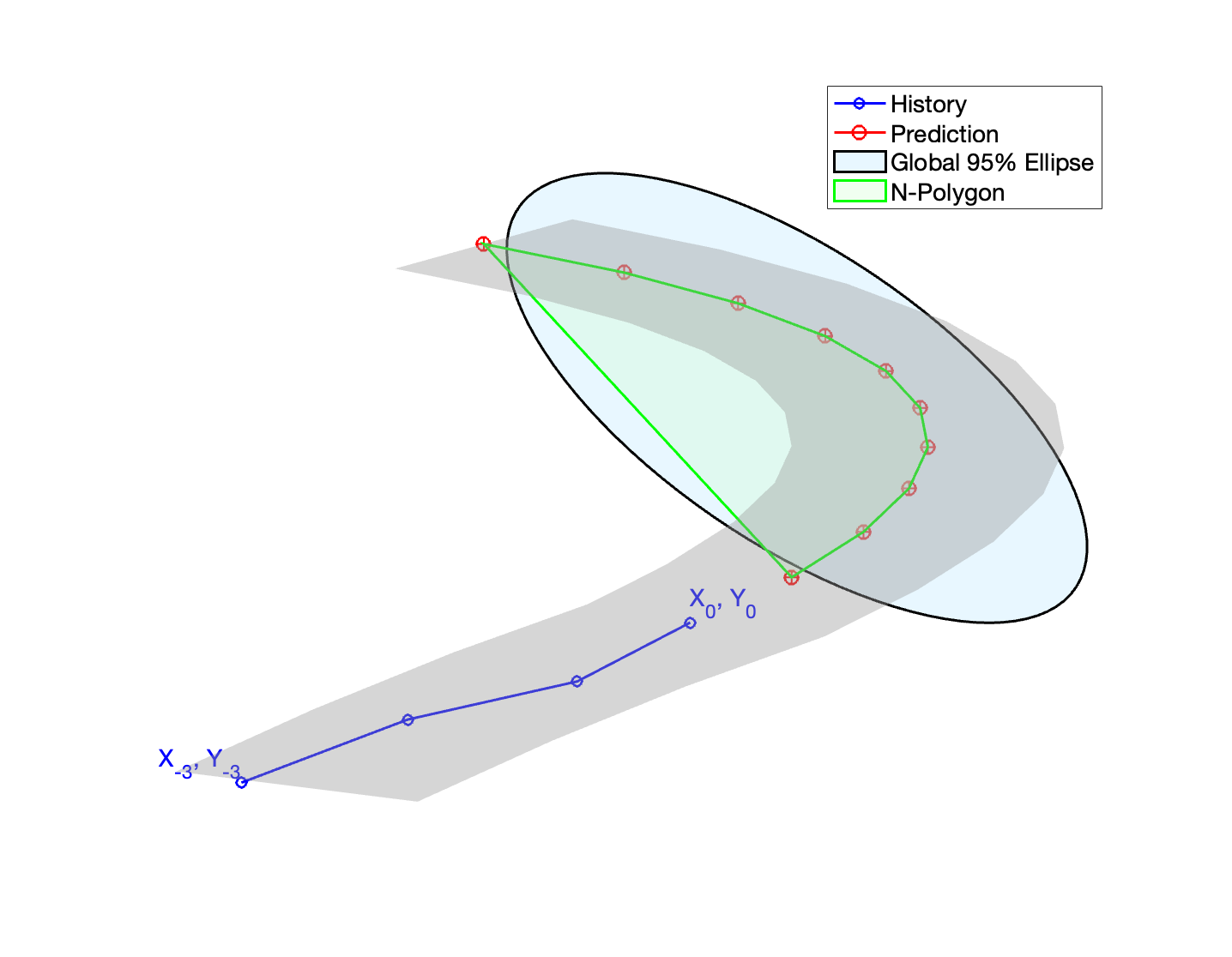}
    \caption{Trajectory Prediction and Representation Forms}
    \label{fig:area-reservation}
\end{figure}

Fig.~\ref{fig:area-reservation} provides a graphical summary of our trajectory prediction process along with the different representation forms considered. It is important to note that the goal of this work is not to develop a high-accuracy prediction algorithm, but rather to explore and evaluate how predicted trajectories can be represented and then communicated efficiently. Accordingly, we employ a simple and lightweight quadratic polynomial fit using the least-squares method (LSM) to estimate the future positions of the \ac{VRU} based on its past motion history. 

Following the prediction, we must select a suitable representation format from those defined in the \ac{ETSI} standard~\cite{etsi-geoarea} to encode the predicted trajectory~\cite{pedestrians}. As illustrated in Fig.~\ref{fig:area-reservation}, three alternative representations are possible: a predicted trajectory vector, consisting of a simple sequence of $T$ predicted $(x, y)$ positions; an uncertainty ellipse, capturing a 95\% confidence region around the prediction; and an $N$-Polygon, which approximates the area covered by the predicted path using a polygonal shape with $N$ vertices. Section~\ref{forms} provides a detailed description of each representation form, while Section~\ref{complexity} presents a comparative analysis in terms of computational and communication complexity.

\section{Intention Sharing and Intention Detection}

\subsection{Scope within the Dynamic Driving Task}
\begin{figure*}[tb!]
    \centering
    \includegraphics[width=0.88\textwidth]{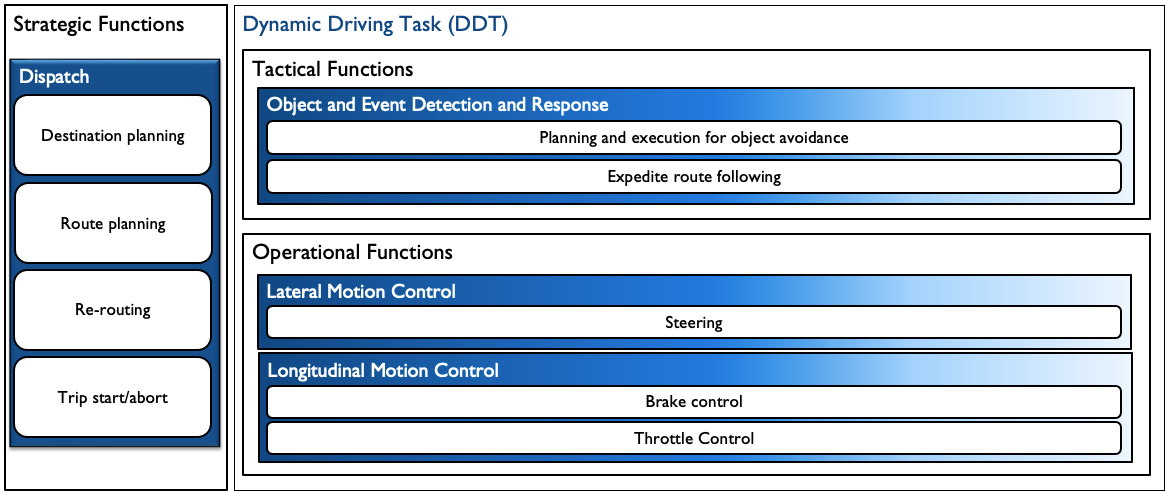}
    \caption{Function, tasks, and sub-tasks in a trip}
    \label{fig:ddt}
\end{figure*}

Fig.~\ref{fig:ddt} shows the division of a trip into Strategic, Tactical, and Operational Functions~\cite{SAELevels}. Strategic Functions include destination and route planning, re-routing, and trip start/abort. Tactical and operational functions are grouped in the \ac{DDT}. Tactical operations include planning and execution for object avoidance and, importantly, expedite route following. Operational functions include longitudinal and lateral movement.

Thus, we place Intention Sharing and Intention Detection close to the \ac{OEDR} sub-task. Natively, even conventional drivers react to the perceived or shared intention of other road users (e.g., when a car tries to merge into a lane and signal that intention to existing traffic). Furthermore, automated vehicles do so using sensors and detection algorithms, and connected also rely on communications to \textit{talk} to each other. Therefore, we study intention sharing and intention detection only as part of the \ac{DDT}.

The implications of this placement are:
\begin{enumerate}
    \item Our time horizon is within the sub-tasks of a maneuver (i.e., in the range of a few seconds).
    \item We consider only small segments of a route that was already planned and we attempt to follow efficiently.
    \item Thus, we share small maneuvers (micro tasks) e.g., overtake, lane switching, 
\end{enumerate}

To illustrate this, consider the example of an overtaking maneuver. This maneuver involves a sequence of lateral and longitudinal control actions: the vehicle first decelerates and changes its heading to merge into an adjacent lane, then accelerates while maintaining lane alignment to pass slower traffic, and finally changes heading again to return to the original lane. Each of these sub-tasks corresponds to a distinct intention that can be communicated via VAMs: (1) the intention to initiate a lane change, (2) the intention to proceed along the new lane while accelerating, and (3) the intention to return to the original lane. In conventional driving, these intentions are implicitly conveyed through the use of turn signals; in connected vehicle systems, they are explicitly shared as structured data (see Fig. \ref{fig:area-reservation}) to increase awareness and safety.

Finally, to further illustrate the relevance and potential of intention sharing in fully cooperative systems, consider the case where the vehicle being overtaken communicates that the adjacent lane is unsafe to merge into (e.g., due to oncoming traffic on a two-lane expressway). This scenario exemplifies maneuver coordination—a concept that extends beyond the traditional scope of intention sharing but uses it as a basis

\subsection{Representation: requirements, and trade-offs} \label{forms}



The choice of which geometric representation to use for VRU trajectories critically impacts communication overhead, computational complexity, and prediction accuracy in V2X systems. We analyze three ETSI accepted representations: predicted trajectory vectors, covariance ellipses, and $N$-point polygons.

\textbf{Trajectory Vector:} The default representation defined in the ETSI standards. This approach encodes a predicted trajectory as an ordered sequence of $T$ future positions, each defined by $(x, y)$ coordinates. It offers a direct and intuitive representation suitable for linear or piecewise-linear motions. Communication cost scales linearly with the prediction horizon, requiring $2T$ floats per trajectory. Computationally, collision checking between trajectories involves segment intersection tests with complexity $\mathcal{O}(T^2)$ per pair, which can become expensive for long horizons or dense networks. Accuracy depends on the sampling rate; higher temporal resolution improves fidelity but increases data volume and computational cost.

\textbf{Covariance Ellipse:} A probabilistic trajectory can also be represented using a Gaussian process, where the uncertainty of the future position is described by a 2-D elliptical covariance matrix. 
This probabilistic representation efficiently captures uncertainty, enhancing robustness to noise and prediction errors. Communication load is significantly smaller than for predicted trajectory vectors due to only requiring the covariance parameters, $5$ floats per trajectory since we are only interested in checking the last prediction step (single ellipse). For collision detection we can leverage ellipse-overlap tests, which scale linearly and are computationally efficient.  However, ellipses approximate uncertainty contours as smooth, convex shapes and may lack the geometric precision needed for complex or non-convex trajectories.

\textbf{$N$-Point Polygon:} This approach approximates the predicted trajectory envelope as a polygon with $V$ vertices, enabling flexible and accurate representation of arbitrary shapes, including non-convex and irregular motion envelopes. Again since we only care about the representation at the final prediction step the communication cost scales as $2 V$, which can be substantial for large values of $V$. Collision checking employs algorithms such as the Separating Axis Theorem, with complexity $\mathcal{O}(V^2)$ per polygon pair per timestep in worst case, though practical optimizations often reduce this cost. This representation offers the highest geometric fidelity but at increased communication and computational expense, making it suitable when precision is paramount and system resources permit.

The choice between these representations depends on application-specific requirements. Trajectory vectors are favorable for simplicity and interpretability but may struggle with uncertainty representation and computational scalability. uncertainty ellipses strike a balance between uncertainty modeling and efficiency but may oversimplify predicted trajectory shapes. $N$-point polygons provide geometric flexibility at the cost of higher resource consumption.

\subsection{Computational complexity} \label{complexity}


This subsection provides a comprehensive analysis of the computational and communication complexities associated with the two most representative V2X coordination paradigms: Intention Sharing (IS) and Intention Detection (ID). In both scenarios, it is assumed that each user predicts either its own future trajectory or those of surrounding agents over a fixed horizon of $T$ time steps, employing a predefined method such as Least Squares Minimization, Kalman Filtering, or Cubic Spline Interpolation.

\subsubsection{Intention Sharing}

In the intention sharing scenario, each user explicitly communicates its intended future trajectory, significantly reducing local inference requirements. As previously mentioned, each user predicts only its own path using a method of complexity $\mathcal{O}(C)$, and transmits the resulting trajectory to neighboring vehicles. The subsequent computational complexity arises primarily from collision checking, which depends on the chosen representation form:

\begin{itemize}
    \item \textit{Trajectory Vector:} Each trajectory contains $T$ predicted positions, resulting in $2T$ floats or $8T$ bytes that need to be communicated per user. If the uncertainty is included, the standard deviation along both the $X$ and $Y$ axes must be transmitted for each point, thereby doubling the overall message size. Collision detection between line segments over $T$ steps requires $\mathcal{O}(T^2)$ comparisons per VRU pair. Given $N$ users, the system-wide complexity becomes $\mathcal{O}(N^2 T^2 C)$.

    \item \textit{Uncertainty ellipse:} The trajectory can be described by a mean position and a $2 \times 2$ covariance matrix, resulting in $5$ floats or $20$ bytes per user. Collision likelihood can be estimated using a center-inclusion heuristic or a ellipsoid-containment test, either procedure involves constant-time operations plus a matrix inversion step of complexity $\mathcal{O}(d^3)$, which reduces to $\mathcal{O}(1)$ for 2D ellipses. Thus, system-wide complexity is $\mathcal{O}(N^2 C)$.

    \item \textit{N-Polygon:} Similar to the ellipse case, the predicted path can be represented using a convex polygon with $V$ vertices, resulting in $4V$ bytes. Collision checking uses the Separating Axis Theorem (SAT), with complexity per polygon pair per timestep approximated as $\mathcal{O}(V)$. The total system-wide complexity is $\mathcal{O}(N^2 C V)$, offering a tunable trade-off between precision and resource usage.
\end{itemize}

Intention sharing shifts the computational burden away from local inference and toward compact, efficient geometric comparisons, benefiting from reduced per-user processing and simplified prediction requirements.

\subsubsection{Intention Detection}

In contrast, the intention detection scenario requires each user to predict the trajectories of $N$ users—including itself—based solely on communicated motion histories over $H$ steps. This leads to a prediction cost of $\mathcal{O}(N T C)$ per user, and a total system-wide cost of $\mathcal{O}(N^2 T C)$. Additionally, collision detection must be performed for all pairs of predicted trajectories, leading to overall complexities of $\mathcal{O}(N^4 T^3 C)$ for predicted trajectory vectors, $\mathcal{O}(N^3 T C)$ for covariance ellipses, and $\mathcal{O}(N^3 T C V)$ for $N$-polygons. 

This approach, while using only motion histories, is computationally intensive since each user must predict trajectories for $N$ neighbors and itself. It is also sensitive to noise and behavioral uncertainty, even before factoring in the complexity of the prediction algorithm and collision checking.

Table~\ref{tab:complexity_analysis} summarizes the computational complexity and communication overhead associated with the different trajectory representation forms. For our proposal, we selected the uncertainty ellipse representation, as it offers a favorable trade-off between low per-message communication overhead and bounded computational complexity.

\begin{table*}[htbp]
\caption{Summary of the Complexity Analysis for Different Representation Forms}
\centering
\begin{tabular}{|l|c|c|c|c|}
\hline
\textbf{Representation Form} & \textbf{ID} & \textbf{IS} & \textbf{Communication Overhead} & \textbf{Uncertainty Representation} \\
\hline
Trajectory Vector & $O(N^4T^3C)$ & $O(N^2T^2C)$ & $8T$ Bytes & Explicit \\
\hline
Uncertainty Ellipse & $O(N^3TC)$ & $O(N^2C)$ & $5$ Bytes & Implicit \\
\hline
N-Polygon & $O(N^3TCV)$ & $O(N^2CV)$ & $V$ Bytes & Implicit \\
\hline
\end{tabular}
\label{tab:complexity_analysis}
\end{table*}

\section{Evaluation}
\subsection{Integration into VAM and influence on network performance and safety}

We simulate \ac{VAM} exchanges using Artery~\cite{Artery}. The scenario contains pedestrians and cyclists going on a segment of a pedestrian/cyclist pathway with crossings on the extremes. We evaluate an ETSI VAM that follows the standardized generation scheme (sending low-frequency containers with a set of 23 coordinates every 2\,s), and another scheme that always sends a container a definition for the shape.

\begin{figure}[tbh!]
    \centering
    \includegraphics[width=\linewidth]{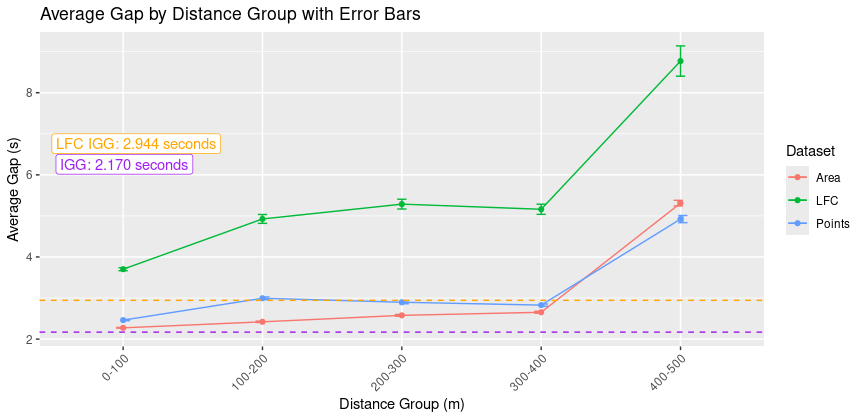}
    \caption{Inter-generation and inter-packet gaps for VAMs with and without low-frequency containers}
    \label{fig:ipg-igg}
\end{figure}

Fig.~\ref{fig:ipg-igg} shows the average Inter-packet Gaps (IPGs) ($t$ between two successful receptions from a node) at different distances compared to how frequently messages are generated (inter-generation Gap --- IGG). The first interesting result is that, in general terms, our version of VAM (i.e., sending an area) surpasses the behavior of the aggregate ETSI VAM (with and without low-frequency container). This is explained by the normalization of VAM sizes (all are of the same length) as opposed to the standard scheme with varying sizes. Furthermore, messages with the low-frequency container are generated every 2.9\,s on average, but are received every 4--5\,s even at distances under 300\,m. This means that the information necessary to share intentions using points is lost at significant rates. Thus, our proposal conveys the same information more effectively and efficiently.

\subsection{Physical implementation}

\begin{figure}[tbh!]
    \centering
    \includegraphics[width=\linewidth]{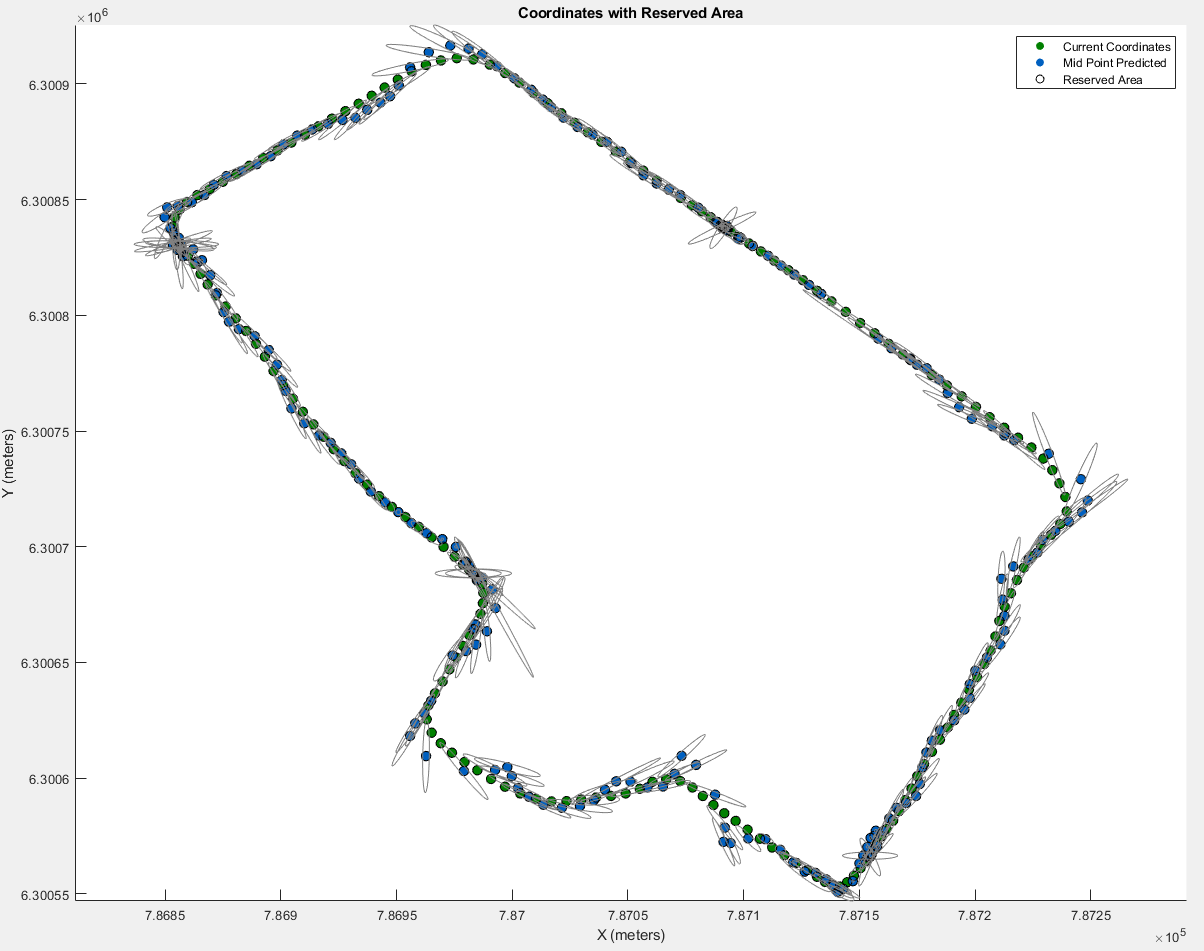}
    \caption{Real-world experiment: coordinates and reserved area}
    \label{fig:real-world}
\end{figure}

Fig.~\ref{fig:real-world} shows the results of our physical implementation on a Raspberry Pi 4 equipped with a USB GNSS sensor. Our experiment showed the feasibility of using embedded devices to run the prediction model. However, low-cost commercial off-the-shelf GNSS sensors have limitations that can be solved, e.g., using Kalman filters.

\section{Discussion and Future Work}

Future efforts will focus on defining the prediction model for VRU trajectories to support accurate and efficient maneuver coordination. Two promising model-based approaches are the Extended Kalman Filter (EKF) and cubic spline interpolation. EKF provides a principled way to account for process and measurement noise, making it well-suited for dynamic state estimation under uncertainty. Cubic splines, on the other hand, offer smooth trajectory generation while preserving continuity and curvature, which are important for realistic bicycle motion modeling.

In contrast, machine learning methods are less suitable for this context due to several constraints. First, the sparsity and variability of bicycle trajectory datasets, especially in the V2X setting, limit generalization. Second, the real-time and distributed nature of V2X communication places stringent requirements on model interpretability, computational efficiency, and predictability—criteria in which black-box models often fall short. Additionally, the need for on-device inference on resource-constrained platforms further reduces the practicality of data-driven approaches.


Furthermore, we intend to leverage intention sharing to enhance the existing ETSI standard clustering mechanism. By enabling intention-based clustering of proximate agents, this approach has the potential to outperform conventional distance-based methods in V2X scenarios. Such improvements could facilitate more scalable and behavior-aware coordination strategies in dense, heterogeneous traffic environments.

\section{Conclusion}
We presented an Intention Sharing adaptation to the VAM protocol to be used by Electrically Power-Assisted Cycles to communicate its predicted trajectory with other V2X-enabled vehicles. We show that representing the predicted trajectory as an ellipse instead of a set of points is more efficient and effective in terms of networking and safety metrics, and that embedded systems have the capability to run physics-based models that can predict motion accurately.

\section*{Acknowledgments}
This work was supported by the ELLIIT network through project 6G-V2X "6G Wireless, sub-project: Vehicular Communications", and the Swedish Transport Administration through the project \textit{''Here I go'' – avancerade funktioner för VRU-medvetenhetsprotokoll i C-ITS} (TF 2023/104170).

\bibliographystyle{ieeetr}
\bibliography{references}


\end{document}